# Echo spectroscopy in multi-level quantum-mechanical rotors

Dina Rosenberg[1,2], Ran Damari[1,2] and Sharly Fleischer[1,2*]

[1]Raymond and Beverly Sackler Faculty of Exact Sciences, School of Chemistry, Tel Aviv University 6997801, Israel.
[2]Tel-Aviv University center for Light-Matter-Interaction, Tel Aviv 6997801, Israel
Email: sharlyf@post.tau.ac.il

Abstract: The rotational echo response of molecules is found to strongly depend on the delay between the two ultrashort laser pulses, as opposed to two-level systems. We study this dependence experimentally and theoretically and show that by judicious control of the 2$^{nd}$ pulse intensity, 'rotational echo spectroscopy' in a multi-level molecular system becomes possible.

---

Echo spectroscopy is a vastly utilized technique in magnetic resonance (MR) spectroscopy [1] and imaging [2], 2D electronic [3–5] and vibrational spectroscopy [6–9], that enables one to experimentally decipher dephasing from decoherence dynamics and determine their rates selectively. Only recently has echo spectroscopy emerged into gas-phase rotational dynamics in a series of works demonstrating alignment [10] and orientation [11] echoes induced by ultrashort optical and terahertz pulses respectively. In an elegant interplay between the inherently periodic rotational dynamics and the induced echo responses, 'fractional echoes' [12] and 'imaginary echoes' [13] were demonstrated and even 'rotated echoes' [14] induced by polarization-skewed pulses. Motivated by utilizing alignment echoes for gas-phase rotational spectroscopy we recently demonstrated the rephasing of centrifugally-distorted molecular rotations via alignment echoes (ALEC) in methyl-iodide [15] and found that while they share the basic physics of two-level photon-echoes, they substantially differ by several other traits discussed hereafter.

In this work we study the dependence of ALEC on the delay between the two excitation pulses. This dependence is absent from two-level systems and emanates from multiple transition pathways that interfere to create rotational coherences within the multi-level rotational system and govern the observed dynamics. We further show that judicious control of the rephasing pulse intensity facilitates multi-level rotational echo spectroscopy and offers additional desirable spectroscopic capabilities.

Coherent rotational dynamics:

Laser-induced molecular rotations has been thoroughly explored for more than three decades [16–19].Since the pioneering works of rotational coherence spectroscopy [20], rotational control became an essential component in various experimental techniques aiming to extract 'molecular frame' spectroscopic signatures (e.g. high-harmonic-generation [21–23], ultrafast X-ray diffraction [24], photoelectron [25] and Coulomb-explosion [26] imaging). In brief, an ultrashort (~100fs duration) laser-pulse imparts torque to molecular rotors, resulting in their rotation toward the pulse polarization direction (z-axis) and their preferred angular distribution along the z-axis (alignment). Throughout their field-free rotation, the rotors dephase and regain their isotropic distribution shortly after. However, due to quantization of angular momentum, the rotational dynamics is inherently periodic and manifest in recurrences of the alignment with each period of the motion, giving rise to a series of alignment events separated by $T_{rev} = (2Bc)^{-1}$, termed revival period ($B$ is the molecular rotational constant in [cm$^{-1}$] and $c$ the speed of light) [27,28].

Echo spectroscopy in two-level systems

A typical photon-echo experiment includes two, time-delayed pulses. The first (with a π/2 area) induces coherent superposition of the two-levels followed by field-free evolution for time Δτ (delay between pulses). During this 'waiting time' (Δτ) the system experiences dephasing and decoherence that manifest by the decay of the signal. At t=Δτ the second pulse (with an area of π) is applied to effectively reverse the time-evolution such that after another period of free-evolution, at t=2Δτ, the system is "in-phase" again and an echo signal is observed [1]. By repeating the experiment and recording the echo as a function of Δτ, one is able to extract the rates of decoherence and dephasing selectively. However, for the abovementioned scheme to be valid, *it is crucial that the echo response is inherently independent of Δτ*, namely that the effective area of the second pulse (optimally π-pulse) is conserved regardless of the delay between pulses. *While this condition is inherently satisfied in two-level systems* (even for pulse areas other than π/2 and π respectively [29]), *it does not hold in multi-level rotational systems that are of interest in this work.*

Experimental setup: The rotational dynamics of carbonyl-sulfide (OCS) was monitored via the weak-field polarization detection technique [10] [30,31]as reported previously [15,32,33]. A 100fs pulse is split to form a pump beam (800nm) and a probe beam (400nm, via BBO crystal) polarized 45$^0$ to the pump. The pump is routed through a Michelson-interferometer with computer-controlled delay stage on one of its arms to yield two collinear pump pulses with controlled intensities and delay apart ($P_1$ and $P_2$). The pumps and probe beams were focused by an f=150mm lens to cross at their Rayleigh range inside the sample cell at a small angle to assure minimal time-averaging across their intersection. We note that the crossed beam geometry limits the interaction volume (<1mm) and thus necessitates higher gas densities in comparison to collinear beams geometry [15,32,33] (few tens of torrs vs. few torrs). As will become clearer later on, this geometry is crucial for our experiment. The diameter of the pump beams is controlled by an iris and their energies selectively varied by two attenuators, and reported hereafter in [mW] such that 1mW corresponds to 2μJ (per pulse). The ALEC response manifest as transient optical birefringence and monitored by the change in the probe polarization [34]. The observed signal is proportional to the change in the degree of alignment $\Delta I/I(t) \propto \left[\langle \cos^2 \theta \rangle_{(t)} - \frac{1}{3}\right]$ where $\theta$ is the angle between the molecular-axis and the z-axis and 1/3 is the isotropic degree of alignment [32,35,36].

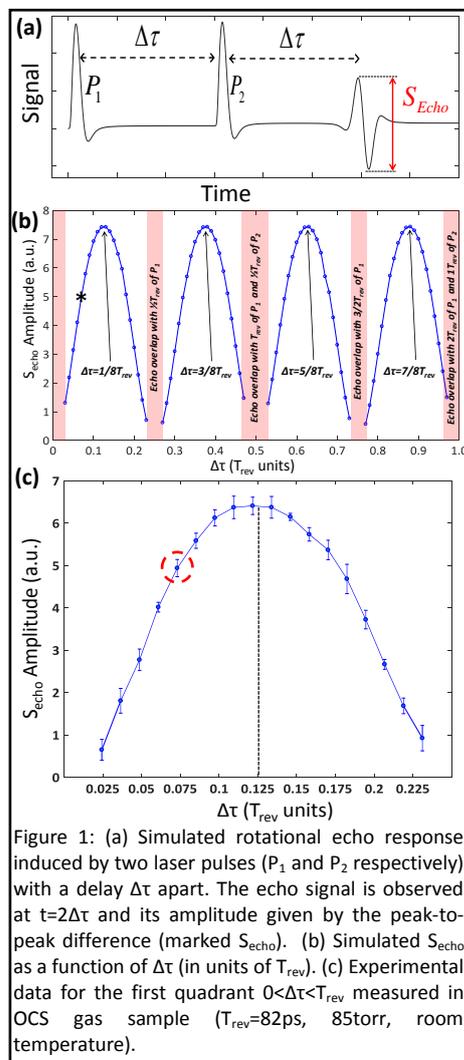

Figure 1: (a) Simulated rotational echo response induced by two laser pulses ($P_1$ and $P_2$ respectively) with a delay Δτ apart. The echo signal is observed at t=2Δτ and its amplitude given by the peak-to-peak difference (marked $S_{echo}$). (b) Simulated $S_{echo}$ as a function of Δτ (in units of $T_{rev}$). (c) Experimental data for the first quadrant 0<Δτ<$T_{rev}$ measured in OCS gas sample ($T_{rev}$=82ps, 85torr, room temperature).

Figure 1 depicts the simulated (Fig.1b) and experimentally measured (Fig.1c) ALEC amplitudes as a function delay between the two pulses(Δτ) with fixed pulse intensities ($P_1$=9.2mW,$P_2$=4.2mW).We quantify the ALEC amplitude by its peak-to-peak difference (marked $S_{echo}$ in Fig.1a) [15].For example, the point marked '∗' in Fig.1b is the echo induced by two pulses with a delay Δτ=0.07$T_{rev}$ apart and its amplitude, $S_{echo}$ observed at t=2Δτ=0.14$T_{rev}$). We find parabolic-like dependence of $S_{echo}$ on Δτ in each quadrant of $T_{rev}$, in excellent agreement with the experimental results of Fig1.c where $S_{echo}$ vs. Δτ at the first quadrant [0<Δτ<0.25$T_{rev}$] were measured in OCS gas (85torr, room temperature). From Figs.1b one sees that the most efficient ALEC response (maximal $S_{echo}$ amplitude) are induced at delays of 1/8$T_{rev}$ (3/8, 5/8, 7/8$T_{rev}$). In what follows we analyze this dependency in the coherent-control framework and show that it results from the interference of multiple quantum-mechanical pathways that lead to the same final rotational coherence. While the different pathways can be represented by double-sided Feynman diagrams [11,15,37], we believe that their interferences are better conveyed by a two-dimensional representation inspired by the rotational density matrix (RDM).

Theoretical model:

Molecules are usually treated as quantum-mechanical rigid rotors in consideration of their rotational dynamics [17]. With $\hat{L}$ as the angular momentum operator, $I$ - moment of inertia, $\Delta\alpha$ - anisotropic polarizability of the molecule, $|E_{(t)}|^2$ - pulse envelope and $\theta$ - the polar angle in spherical coordinates, the Hamiltonian is given by $\hat{H} = \hat{L}^2/2I - \frac{1}{4}\Delta\alpha|E_{(t)}|^2(t)\cos^2\theta$ [ref]. Our numerical simulations are performed by propagating the Liouville-von-Neumann equation $\partial\rho/\partial t = -\frac{i}{\hbar}[\hat{H},\rho]$ using the density matrix formalism described previously [15]. The eigenstates of the rotors are the spherical harmonic functions $|J,m\rangle$ and quantized in the $\hat{z}$-axis direction, the polarization direction of the two interacting pulses for convenience. Since $\hat{H}$ is independent of the azimuthal angle $\varphi$, it invokes the selection rule of $\Delta m = 0$ for transitions allowing us to restrict the discussion to the $J$ quantum-number solely. The interaction operator, $\hat{V} \propto \cos^2\theta$, summons upon transitions with $\Delta J = \pm 2$ only. We consider only ALEC responses induced via one and two Raman interactions with the first ($P_1$) and second ($P_2$) pulses respectively ($S_{echo} \propto P_1,(P_2)^2$ as shown in [15]). The $P_1$-induced transition is depicted by straight arrow and $P_2$-induced transitions by (two) curved arrows in Fig.2)

Before its first interaction, the system's thermal populations represented by the dark-red dots along the diagonal of the RDMs (the pale-red terms do not participate in the specific scenarios that are presented).

The 3x3 RDM in Fig.2a is an effective two-level system since both $P_1$ and $P_2$ induce Raman transitions with $\Delta J = \pm 2$ that do not couple the $J+1$ state to its neighboring states. $P_1$(black arrow) induces rotational coherence $|J\rangle\langle J+2|$ (blue dot, 2$^{nd}$ off-diagonal) that governs the molecular alignment and its dynamics [37]. The phase accumulated by this term for time $\Delta\tau$ is:

$e^{-i\phi_{|J\rangle\langle J+2|}} = \exp[-i(E_{J+2} - E_J)\Delta\tau/\hbar]$ (with $E_J = BJ(J+1)$ and $B = \hbar^2/2I$ for rigid-rotors). $P_2$ is applied at $t = \Delta\tau$ and interacts with $|J\rangle\langle J+2|$ via two Raman transitions to create the conjugate coherence term $|J+2\rangle\langle J|$, the latter

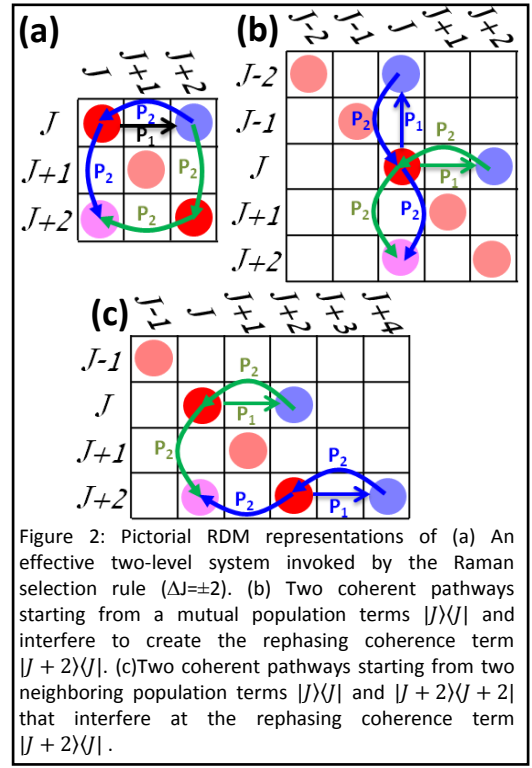

Figure 2: Pictorial RDM representations of (a) An effective two-level system invoked by the Raman selection rule ($\Delta J=\pm 2$). (b) Two coherent pathways starting from a mutual population terms $|J\rangle\langle J|$ and interfere to create the rephasing coherence term $|J+2\rangle\langle J|$. (c) Two coherent pathways starting from two neighboring population terms $|J\rangle\langle J|$ and $|J+2\rangle\langle J+2|$ that interfere at the rephasing coherence term $|J+2\rangle\langle J|$.

accumulates phase at the exact same frequency only with a negative sign $e^{-i\phi_{|J+2\rangle\langle J|}} = \exp[-i(E_J - E_{J+2})\Delta\tau/\hbar]$. At $t = 2\Delta\tau$ the total accumulated phase is $\phi_{|J\rangle\langle J+2|} + \phi_{|J+2\rangle\langle J|} = 0$, and a fully rephased echo response is observed. In fact, even for the two-level case, there are two interfering pathways (blue and green curved

arrows). However since both pathways share the $|J\rangle\langle J+2|$, they accumulate the exact same phase $\exp(-i\phi_{|J\rangle\langle J+2|})$ and their phase difference remains zero for all $\Delta\tau$'s and their interference at the final $|J+2\rangle\langle J|$ term and the corresponding rephased echo amplitude is independent of $\Delta\tau$.

In our multi-level rotational system, P$_1$ and P$_2$ induce multiple interfering pathways exceeding beyond the $2\times 2$ space spanned by two-level systems. Those different pathways accumulate phase difference as a function of the 'waiting time' $\Delta\tau$ and result in the observed ALEC dependence on $\Delta\tau$ (Fig.1b,c). Figs.2b,c are exemplary cases where different pathways start at a mutual initial population term or at adjacent terms respectively.

Case 1: Single population term (Fig.2b) - Consider $|J\rangle\langle J|$ as the initial population term. P$_1$ provides one Raman interaction to create both the $|J-2\rangle\langle J|$ and $|J\rangle\langle J+2|$ (blue and green P1 arrows) that accumulate phase as $\exp(-i\phi_{|J-2\rangle\langle J|}) = \exp[-i(E_J - E_{J-2})\Delta\tau/\hbar]$ and $\exp(-i\phi_{|J\rangle\langle J+2|}) = \exp[-i(E_{J+2} - E_J)\Delta\tau/\hbar]$ respectively. How those two pathways interfere at the final rephasing term depends on the accumulated phase difference at the time of interaction with P$_2$
$$\Delta\phi = \phi_{|J\rangle\langle J+2|} - \phi_{|J-2\rangle\langle J|} = [(E_{J+2} - E_J) - (E_J - E_{J-2})]\Delta\tau/\hbar = 8B\Delta\tau/\hbar \quad (1).$$

Thus, by setting $\Delta\tau = h/16B$ or $\tfrac{1}{8}T_{rev}$ ($\tfrac{3}{8}, \tfrac{5}{8}, \tfrac{7}{8}T_{rev}$) in units of the rotational revival period ($T_{rev} = h/2B$), the two pathways accumulate a $\pi$ ($3\pi, 5\pi, 7\pi$ respectively) - phase apart and *constructively interfere* to create the $|J+2\rangle\langle J|$ maximal rephasing coherence and the largest S$_{echo}$ as shown in Fig.1. In accordance, for $\Delta\tau = \tfrac{2}{8}, \tfrac{4}{8}, \tfrac{6}{8}T_{rev}$ the two pathways accumulate a $2\pi, 4\pi, 6\pi$ phase difference respectively that lead to minimal rephasing coherence (and minimal S$_{echo}$) due to their destructive interference at $|J+2\rangle\langle J|$. In conclusion, the two coherent pathways are inherently destructively interfering, unless a phase (optimally $\pi$-phase) is introduced by setting $\Delta\tau$ to $\tfrac{2n+1}{8}T_{rev}$ ($n \in \mathbb{N}$).

Case 2: Two adjacent population terms (Fig.2c) - In a thermally populated rotational system, interfering pathways may also originate from different population terms $|J\rangle\langle J|$ and $|J+2\rangle\langle J+2|$ as exemplified in Fig.2c. While the two pathways marked by the green and blue arrows start as incoherent, their induced transitions interfere at the $|J+2\rangle\langle J|$ coherence (pink dot). Also here, the interference depends on the phase difference between the two pathways which accumulates with $\Delta\tau$ to
$$\Delta\phi = \phi_{|J\rangle\langle J+2|} - \phi_{|J+2\rangle\langle J+4|} = 8B\Delta\tau/\hbar.$$

While Fig.2 and associated text consider the lowest number of interactions that govern the echo response, we note that higher orders of interaction (e.g. two and three interactions with $P_1$ and $P_2$ respectively) may also contribute to the observed ALEC and are accounted for in our simulations [34], and their ramifications discussed hereafter.

ALEC dependence on both Δτ and $P_2$

As $P_1$ and $P_2$ intensities increase, higher orders of interaction gradually alter the dependence of $S_{echo}$ on the pump pulses' intensities. While the linear dependence of $S_{echo}$ on $P_1$ is retained, its quadratic dependence on $P_2$ evolves into an oscillatory dependence, the initial (low $P_2$ intensities) region of which can be fitted by a sinusoidal squared as reported previously [15]. With even higher $P_2$ intensities, $S_{echo}$ is found to oscillate and gradually decay.

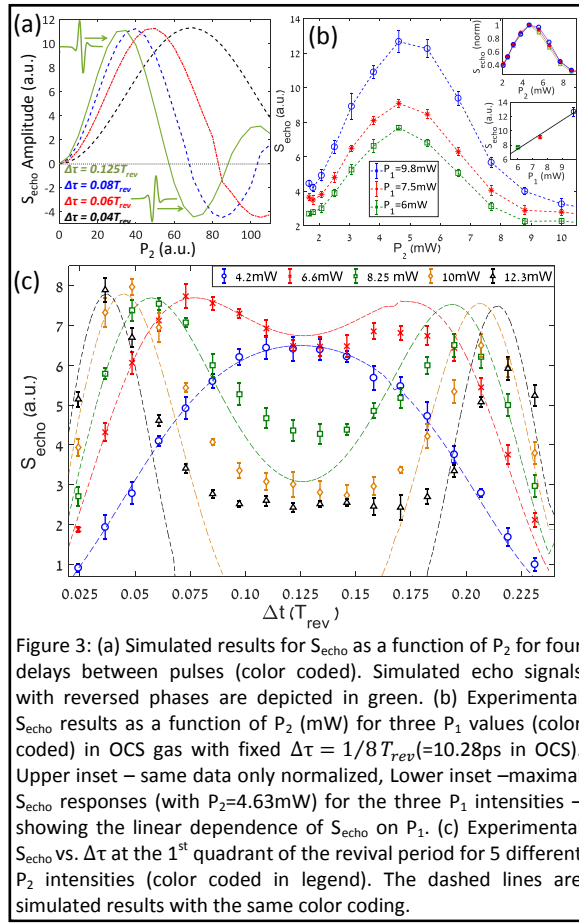

Figure 3: (a) Simulated results for $S_{echo}$ as a function of $P_2$ for four delays between pulses (color coded). Simulated echo signals with reversed phases are depicted in green. (b) Experimental $S_{echo}$ results as a function of $P_2$ (mW) for three $P_1$ values (color coded) in OCS gas with fixed $\Delta\tau = 1/8\, T_{rev}$ (=10.28ps in OCS). Upper inset – same data only normalized, Lower inset –maximal $S_{echo}$ responses (with $P_2$=4.63mW) for the three $P_1$ intensities – showing the linear dependence of $S_{echo}$ on $P_1$. (c) Experimental $S_{echo}$ vs. Δτ at the 1st quadrant of the revival period for 5 different $P_2$ intensities (color coded in legend). The dashed lines are simulated results with the same color coding.

Fig.3a depicts simulation results of $S_{echo}$ vs. $P_2$ for different delays Δτ=(0.04, 0.06, 0.08, 0.125$T_{rev}$), showing a decaying oscillatory behavior of $S_{echo}$ as $P_2$ increases. Note that negative $S_{echo}$ values correspond to phase-inversed ALEC exemplified in Fig.3a (green ALEC transients). Additional simulations performed for an extended range of $P_2$ intensities revealed that the shape of $S_{echo}$ vs. $P_2$ is conserved for all Δτ's namely, by simple linear rescaling of the $P_2$-axis all of the curves (differing by Δτ between pulses) may collapse onto one 'master curve'. The phase inversion of $S_{echo}$ serves as an indication for higher-orders of interaction that kick-in at increased $P_2$ intensities and are left beyond the scope of this work. Thus, we restrict our discussion to the lower $P_2$ region, where $S_{echo}$ can be fitted by $S_{echo} = a \cdot \sin^2(b \cdot P_2)$ and where strong-field effects such as ionization are experimentally avoided. Fig.3b shows experimental results of $S_{echo}$ vs. $P_2$ with fixed $\Delta\tau = 1/8\, T_{rev}$ for three $P_1$ values: 6,7.5,9.8mW (12,15,19.6uJ/pulse respectively). The insets depict the normalized experimental data and the linear dependence [15] of $S_{echo}$ on $P_1$, verifying that *the modulation of $S_{echo}$ with $P_2$ is fully decoupled from $P_1$ intensity*. Fig.3c depicts the experimentally measured $S_{echo}$ vs. Δτ (in the range $0 < \Delta\tau < 1/4\, T_{rev}$) for a fixed $P_1$=9.2mW and five different $P_2$ values (4.2mW [blue], 6.6mW [red], 8.25mW [green], 10mW [yellow] and 12.3mW [black]). Different $P_2$ intensities yield significantly different $S_{echo}$ progressions as readily observed in the figure. For low $P_2$ intensities ($P_2 \leq 4.2 mW$ blue data), $S_{echo}$ peaks at

$\Delta \tau = 1/8\, T_{rev}$ with parabolic-like progression similar to Figs. 1b,c. For $P_2 \geq 6.2 mW$ (other four curves), the maximal $S_{echo}$ amplitudes are found at $\Delta \tau$'s other than $1/8 T_{rev}$. This is consistent with the oscillatory $S_{echo}$ dependence on $P_2$ shown in Fig.3a where for $P_2<30$ (in the arb.u. of the simulation), the maximal $S_{echo}$ amplitude is always found at $\Delta \tau = \tfrac{1}{8} T_{rev}$ while for $P_2>30$ the $\Delta \tau$ for which maximal $S_{echo}$ is induced gradually shifts to $\Delta \tau < \tfrac{1}{8} T_{rev}$ and to $\Delta \tau > \tfrac{1}{8} T_{rev}$ with a fairly symmetric progression. The dashed lines are simulated results (color-coded) and seem to capture the experimental trends well, however much better agreement is found for the low $P_2$ intensities (blue, red and green) than for the high ones (orange, black). In addition to decoherence effects (mostly collisions, not included in our simulations), the main reason for discrepancies between the simulated and experimental results is attributed to the inevitable averaging over the (Gaussian) intensity distribution of the pump pulses by the probe beam. It is the broad range of $P_2$ intensities experienced by molecules positioned at different locations within the interaction volume that results in a range of echo amplitudes and even phase reversals as shown in Fig.3a) – all averaged over by the probe beam. As $P_2$ increases, so does the range of its intensities and the corresponding averaging that leads to larger discrepancy (orange and black data sets). While the abovementioned averaging can be partially reduced by the crossed-beams geometry, the fine-structure of ALEC within the interaction volume remains mostly hindered.

To this point, we have studied the intricate dependence of the rotational echo response on $P_1$, $P_2$ and $\Delta \tau$. Unlike two-level systems where those three "experimental knobs" are fully decoupled, in our multi-level system $P_2$ and $\Delta \tau$ are interweaved via multiple interference pathways (Fig.2) and result in the convoluted rotational echo response shown in Fig.3. *The cross-dependence of $S_{echo}$ on both $P_2$ and $\Delta \tau$ obstructs conventional applications of echo-spectroscopy in multi-level rotational systems, such as the selective characterization of decay and decoherence dynamics –a key feature of echo spectroscopy.* One possibility is to restrict the examination of $S_{echo}$ to $\Delta \tau$'s that are synced with the revival-period ($\Delta \tau + nT_{rev}$, $n \in \mathbb{N}$), however clearly limited by the decay and decoherence rates, i.e. to samples of sufficiently low densities. Comparing to numerical simulation results like those of Figs.1,3 is yet another option, but requires exact knowledge of the system's dynamics such as the dependence of collisions on the J,M quantum numbers [38].

In what follows we propose and demonstrate a procedure that overcomes the abovementioned restrictions and enables rotational echo spectroscopy. The strategy relies on the finding that *the maximal obtainable ALEC ($S_{echo}^{\max}$) is dictated solely by $P_1$* (and gas parameters – B, $\Delta\alpha$, temperature). This is readily observed in Fig.3a where the maximal $S_{echo}$ amplitudes of all four curves (differing by their $\Delta \tau$) peak at the same value (~11 in the arb.u. of the simulation). The latter was confirmed by additional decay- and decoherence-free simulations that yielded the same $S_{echo}^{\max}$ with up to 1-2% variation across the $0.02 T_{rev} < \Delta \tau < 0.23 T_{rev}$ region. Thus, $P_1$ imparts coherences that are partially rephased by $P_2$, but the degree to which they are rephasable ($S_{echo}^{\max}$) is independent of the delay $\Delta \tau$ by properly choosing $P_2$ intensity as demonstrated hereafter.

Figure 4 depicts our experimental (green points, blue circles) and simulated (dashed green and blue curves) results of $S_{echo}^{max}$ vs. $\Delta\tau$ with fixed $P_1$ (9.2 mW). For each delay we've monitored the $S_{echo}$ amplitudes induced by a range of $P_2$ intensities and recorded $S_{echo}^{max}$ and the intensity of $P_2$ that yielded it (green points, termed $P_2^{opt}$). We find a minimal $P_2^{opt}$ intensity at $\Delta\tau = T_{rev}/8$ - consistent with the analysis of Fig.2 and in agreement with the simulation results (green dashed curve). The experimental $S_{echo}^{max}$ amplitudes (blue circles) show gradual decay with $\Delta\tau$ due to collisional-dephasing of the OCS gas ensemble (85torr, room-temperature). Thus by fitting the experimental $S_{echo}^{max}$ to an exponent, one can extract the collisional decoherence of the gas selectively. The dashed blue curve depicts our simulated results, multiplied by a decaying exponent to fit the experimental results [39].

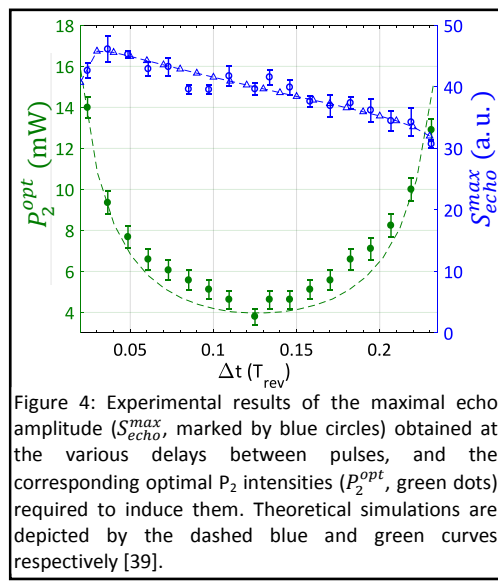

Figure 4: Experimental results of the maximal echo amplitude ($S_{echo}^{max}$, marked by blue circles) obtained at the various delays between pulses, and the corresponding optimal $P_2$ intensities ($P_2^{opt}$, green dots) required to induce them. Theoretical simulations are depicted by the dashed blue and green curves respectively [39].

## Discussion

We have shown that multi-level rotational systems invoke intricate dynamics that arise from interference among multiple quantum pathways. For rotational-echo spectroscopy, those interferences manifest by inherent coupling of the delay between pulses and the intensity of the second pulse. We have found that for each $\Delta\tau$, there is a $P_2$ intensity for which a maximal echo response is induced. Moreover, the amplitude of $S_{echo}^{max}$ is independent of $\Delta\tau$, providing an inclusive "experimental anchor" that is easily extracted by varying $P_2$ intensity and monitoring $S_{echo}$ for a fixed delay between pulses. Then, by varying $\Delta\tau$ and recording $S_{echo}^{max}$, one monitors the decay of $S_{echo}^{max}$ and extracts the desirable decoherence rate. In this work we deliberately experimented with gas samples at low density in order to minimize the collision rates and retain the focus on the basic physics of rotational echoes. However, the strength of the described method is in its applicability to dense gas ensembles. The dynamics of such ensembles, governed by the (high) collision rates are experimentally inaccessible due to coherence life-times that may be shorter than the revival period, much like in liquid phase. The proposed scheme is practically decoupled from the revival period therefore applicable to high-density ensembles. The ability for experimentally identifying the 1/8th revival period (Fig.4) provides yet another advantage with specific implications to large molecules with very long $T_{rev}$ and are underway in our lab. On a broader perspective, our findings can be back-traced to the energy level structure of the quantum-mechanical rotor. It is the quadratic energy dependence on $J$ that dictates the harmonic level-spacing and corresponding periodicity of the rotational dynamics (and the ALEC response). The mathematical analogy between rotational ($\propto J^2$) and anharmonic vibrational ($\propto v^2$) level energies suggest that traits like those described above are expected to manifest in the vastly explored field of vibrational echo spectroscopy as well.

We acknowledge Prof. Jean-Michel Hartmann and Prof. Olivier Faucher for their important remarks and the support of the Israel Science Foundation (ISF) grant no.1065/14 and 926/18, ISF Grant No. 2797/11 (INREP - Israel National Research Center for Electrochemical Propulsion) and the Wolfson Foundation Grants PR/ec/20419 and PR/eh/21797.

* Corresponding author: sharlyf@post.tau.ac.il

# Supplementary Information-

## SI.1 - Experimental setup - Time resolved optical birefringence (Alignment)

Optical-induced alignment of the gas was measured using a time-resolved optical birefringence setup [15], utilizing the 'weak field polarization detection' [30,31].

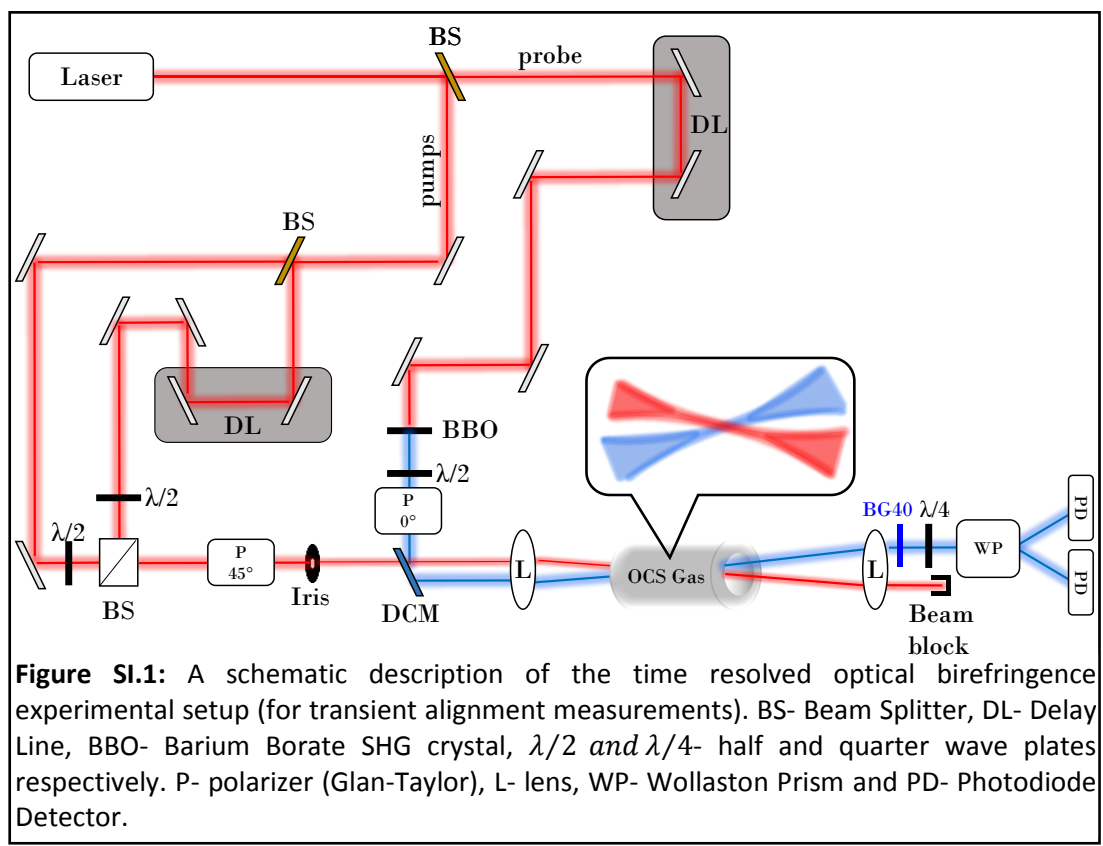

**Figure SI.1:** A schematic description of the time resolved optical birefringence experimental setup (for transient alignment measurements). BS- Beam Splitter, DL- Delay Line, BBO- Barium Borate SHG crystal, $\lambda/2$ and $\lambda/4$- half and quarter wave plates respectively. P- polarizer (Glan-Taylor), L- lens, WP- Wollaston Prism and PD- Photodiode Detector.

An ultrashort laser beam (~100 fs, 800nm) from a Ti:Sapphire Chirped pulse amplifier (Legend-Duo, Coherent Inc.) is split to form a pump (90%) and probe (10%) beams. The pump beam is split by a 50:50 beam splitter to for two pulses with a controlled delay apart and recombined again by a non-polarizing beam-splitter cube to propagate collinearly. The intensities of the two pump beams are selectively controlled by a pair of λ/2 plates (one of which is mounted on a computer controlled rotation stage (with ±0.02° accuracy). The probe beam is frequency doubled in a BBO crystal to form a weak 400nm probe pulse and is delayed by a computer controlled delay stage (DL). The polarization of the 800nm pump beams is set at 45°with respect to the (400nm) probe and propagate parallel to each other after the dichroic mirror (DCM). The beams are focused by a f=150mm lens (L) to cross inside the static gas cell (OCS, 85torr, ambient temperature). Special attention is given for crossing the beams at their Rayleigh range in order to minimize (yet, only partially as mentioned in the main text) the inherent averaging over the intensity span of the pump beam. At the output of the cell, the 800nm pump beam is blocked and its residual scattering is filtered out by a short-pass filter (BG40). The transmitted 400nm probe is re-collimated by a lens and its polarization changes (owing to interaction with the transiently birefringent gas sample) are analyzed via differential polarization detection [35,36]. The latter is achieved by a λ/4 plate, a Wollaston prism (WP) and a pair of balanced photodiode detectors (PD) for the two perpendicular polarizations.

## SI.2– Simulation details

All of the theoretical results presented in this work were extracted from time-dependent rotational dynamics simulations performed by numerically propagating the Liouville-Von-Neumann equation using the density matrix formalism:

$\frac{\partial \hat{\rho}}{\partial t} = -\frac{i}{\hbar}[\hat{H}, \hat{\rho}]$. As done in many previous papers, also here we treat linear molecules as quantum-mechanical rigid rotors [17].

The Hamiltonian of the system, $\hat{H} = \frac{\hat{L}^2}{2I} + V(\theta, t)$, consists of a field-free term, $\frac{\hat{L}^2}{2I}$, with $\hat{L}$ the angular momentum operator and $I$ - the molecular moment of inertia and an interaction term: $V(\theta, t) = -\frac{1}{4}\Delta\alpha |\epsilon|^2(t)\cos^2\theta$. $V(\theta, t)$ is the laser-molecule interaction potential, where the field couples to the molecular rotations via the molecular polarizability tensor. $|\epsilon|^2(t)$ is the Gaussian intensity envelope of the pulse and $\Delta\alpha$ is the anisotropic polarizability of the molecules ($\Delta\alpha = \alpha_\parallel - \alpha_\perp$, parallel and perpendicular to the molecular axis).

The main observable of interest in this work is $\tilde{A} = \langle\langle\cos^2\theta\rangle\rangle$, i.e. the degree of molecular alignment that is induced by two, time delayed, short pulses.

The simulations start with a thermal ensemble (only population terms along the diagonal of the density matrix $\rho$) with $\rho_{JJ} = \exp\left[-\frac{E_J}{k_B T}\right]/Z$ (where Z is the distribution function, $k_B$ and $T$ are the Boltzmann constant and temperature respectively and $E_J = hBcJ(J+1)$, $B = \frac{h}{8\pi^2 cI}$ (the rotational constant). Since both the interaction term $V(\theta, t)$ and the field-free propagator do not mix rotational states with different m quantum numbers, we perform the simulation for the entire J-state manifold but for each m-number at a time (and average them with appropriate weights to obtain the result). This approach significantly reduces the computational cost and reduces the run time dramatically, hence enabling us to simulate the rotational responses of linear molecules at ambient temperatures and over a wide range of experimental parameters (such as the delays between pulses, their varying intensities etc.) on a desktop machine.